\documentclass[amsmath,twocolumn,amssymb,prd,preprintnumbers,showpacs]{revtex4}
\usepackage{amsmath}
\usepackage[cmtip,arrow]{xy}
\usepackage{pb-diagram,pb-xy}
\usepackage{slashed}
\usepackage{amssymb}
\usepackage{slashed}
\usepackage{amssymb,amsopn}
\usepackage{graphicx}
\newcommand{\sla}[1]{\hbox{{$#1$}\llap{$/$}}}
\def \be {\begin{equation}}
\def \ee {\end{equation}}
\def \bea {\begin{eqnarray}}
\def \eea {\end{eqnarray}}
\usepackage{graphicx}
\usepackage{amssymb}
\begin{document}
\title{Unitarity in a Lorentz symmetry
 breaking model with higher-order operators}
\author{Markos Maniatis}
\email[Electronic mail: ]{ mmaniatis@ubiobio.cl}
\author{Carlos M. Reyes}
\email[Electronic mail: ]{creyes@ubiobio.cl}
\affiliation{Departamento de Ciencias 
B{\'a}sicas, Universidad del B{\'i}o B{\'i}o, Casilla 447, 
Chill\'an, Chile.}
\begin{abstract}
We analyze the unitarity of a modified QED with higher-order terms
that violate Lorentz symmetry. We make an explicit calculation 
to verify unitarity at the one-loop level. 
As expected we find negative norm-states that could in principle lead to a 
violation of unitarity. 
However, we show that these states become massive 
in Euclidean space 
and do not contribute to the discontinuity.
\end{abstract}
\maketitle
\section{Introduction}
At energy scales at the Planck mass, $m_\text{Pl} \approx 10^{19}~\text{GeV}$, 
gravitational forces become as strong as the electroweak forces.
We expect that at experiments at the electroweak scale, given 
by the Z-boson mass, $m_Z \approx 91~\text{GeV}$, gravitational effects are suppressed
by a factor $m_Z/m_\text{Pl} \approx 10^{-17}$. 
Nevertheless, gravitational effects of elementary particles 
could be observable at currently available energy scales~\cite{Foam}.
One interesting class of these Planck-mass suppressed effects violate
Lorentz invariance. This possibility has been widely explored in several contexts,
such as in modified dispersion relations~\cite{MDR}, loop 
quantum gravity~\cite{LQG}, and string theory~\cite{Strings,Strings1}. 

In an effective approach, Lorentz-invariance-violating
terms may be imposed at the Lagrangian level.
In Ref.~\cite{SME} the possible Lorentz-violating and renormalizable operators 
were classified. The principal idea is to impose gauge-invariant 
terms, involving
non-Lorentz-invariant tensors with Planck-mass suppressed couplings.
In this way, Lorentz invariance is explicitly broken, but the effective terms 
may originate from a spontaneously broken gravitational theory~\cite{SLSB}.  

Recently the same idea has been extended to consider 
higher-order operators~\cite{Myers-Pospelov,higher-derivative2,SME-nonminimal}.
They also have been considered to solve the 
hierarchy problem in the standard model~\cite{Higgs},
regularization~\cite{Regulator}, and radiative corrections~\cite{Radiative,Reyes,Mariz}.
In particular, in Ref.~\cite{Myers-Pospelov} the modifications
of the dispersion relations originating from the modified kinetic
 terms of the
Lagrangian with respect to scalars, fermions, and vector particles have been
studied. From a comparison to astrophysical experiments, limits on the
new couplings have been derived~\cite{MP_Limits}.

A possible drawback of the higher-order modifications of the kinetic terms is, that 
negative-norm states can appear~\cite{PU,Lee-Wick}. These states correspond to
massive modes of the photon field and could in principle violate unitarity. 
Therefore it appears quite essential to 
verify explicitly that unitarity is preserved~\cite{Schrek}. 
In previous works, it has been shown
for Lorentz-violating modifications of photons~\cite{Unitarity} 
that in Bhabha scattering at the tree level unitarity is fulfilled.
Here we want to proceed and show that 
unitarity is conserved in modified quantum 
electrodynamics at the one-loop level.
\section{The modified photon sector} \label{secphoton}
We will consider the modified QED Lagrangian~\cite{Myers-Pospelov}
\begin{equation} \label{M-M-P} 
\mathcal L=\bar \psi (\sla{\partial} -m)
\psi-\frac{1}{4}F^{\mu\nu}  F_{\mu\nu}- \frac{\xi}{2m_\text{Pl}} 
n_{\mu}\epsilon^{\mu\nu \lambda \sigma} A_{\nu}(n 
\cdot \partial)^2   F_{\lambda \sigma}\;,
\end{equation}
where $n$ is a noninvariant four-vector defining a preferred 
reference frame, $m_\text{Pl}$ is the Planck mass, 
and $\xi$ is a dimensionless coupling parameter. 
Obviously the modification term is a gauge-invariant 
dimension-five operator which is Planck-mass suppressed.

The equations of motion derived from the Lagrangian~\eqref{M-M-P} read
\begin{equation}\label{eqmot}
\partial_{\mu} F^{\mu \nu}+g \epsilon ^{\nu \alpha \lambda \sigma} n_{\alpha} 
 (n \cdot \partial)^2   F_{\lambda \sigma}  =4\pi j^{\nu}\;,
\end{equation}
where we have introduced a source $j^{\nu}$ and defined $g=\xi/m_\text{Pl}$.
In addition we have to impose a gauge fixing, where we 
choose the axial gauge~\cite{Andrianov},
\begin{equation}
\partial_\mu A^\mu  = 0\;, \quad n_\mu A^\mu  = 0 \;.
\end{equation}
\begin{figure*}
\centering
\includegraphics[width=0.97 \textwidth]{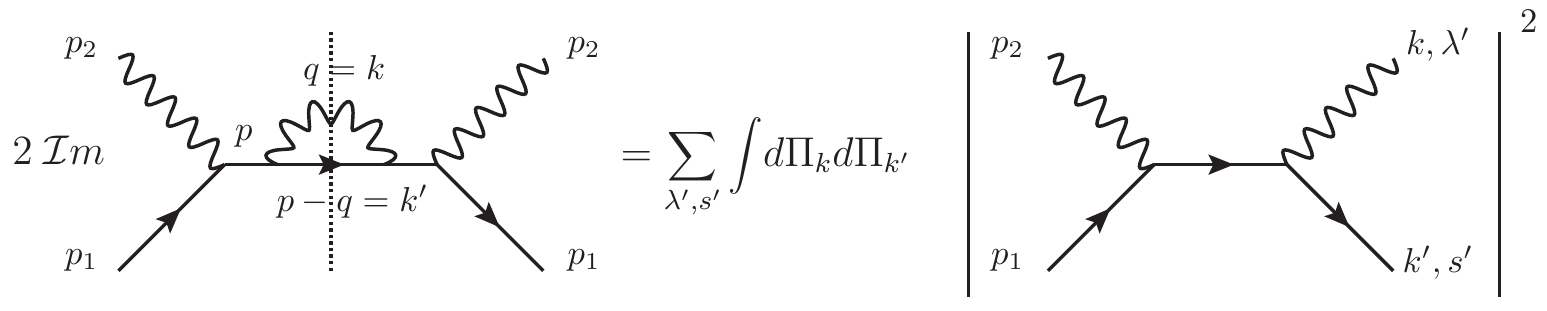}
\caption{\label{Fig1}The optical theorem for a one-loop forward 
Compton scattering amplitude. The integrals
$\int d\Pi_k d\Pi_{k'}$ denote the phase-space 
integration over the final-state momenta.}
\end{figure*}%
In order to verify unitarity, we shall first derive
the modified photon propagator originating from Eq.~\eqref{M-M-P}.
To this end we write the equation of motion in the form
\begin{equation}\label{eq-mot}
M^{\nu \sigma} A_{\sigma}(x)=4\pi j^{\nu},
\end{equation}
where we have defined the operator $M^{\nu \sigma}=\eta^{\nu \sigma}  
\Box+2g \epsilon^{\nu \mu 
\alpha \sigma}n_{\mu} (n\cdot \partial)^2 \partial_{\alpha}$.
In order to construct the modified polarization vectors of the
photon field, let us, following 
closely Ref.~\cite{Andrianov}, introduce the projectors
onto the plane, transverse to both directions, $k$ and $n$,
\begin{multline}
e^{\mu\nu}=\eta^{\mu \nu}-\frac{(n\cdot k)}{D}
(n^{\mu}k^{\nu}+n^{\nu}k^{\mu}) +\frac{k^2}{D} n^{\mu}n^{\nu}
\\
+\frac{n^2}{D} k^{\mu}k^{\nu}\;,
\end{multline}
with $D=(n\cdot k)^2-n^2k^2$.
We easily verify that indeed $k_\mu e^{\mu\nu} = 
n_\mu e^{\mu\nu} =0$ as well as
$e^\mu_\alpha e^{\alpha\nu} = e^{\mu\nu}$.
By means of these projectors we can construct two 
linear polarization vectors $e^{(1)}$ and
$e^{(2)}$ respecting
\begin{equation}
e_{\mu \nu} = - \sum_{a=1,2} e_\mu^{(a)} e_\nu^{(a)} \;,
\quad
\eta^{\mu\nu} e_\mu^{(a)} e_\nu^{(b)} = - \delta^{ab}\;.
\end{equation}
By these linear polarization vectors, modified circular polarization 
vectors are constructed,
\begin{equation} \label{circular}
\varepsilon_\mu(k,\lambda) = \frac{1}{2} \big( e_\mu^{(1)} 
+\lambda i\; e_\mu^{(2)} \big)\;,\\
\end{equation}
with $\lambda =\pm1$ corresponding to left and right polarizations, respectively.
We can see, that the tensor
\begin{equation}\label{imp-rel}
 P^{(\lambda)}_{\mu \nu}=- \varepsilon_{\mu}(k,\lambda)
\varepsilon^{*}_{\nu}(k,\lambda)\;,
\end{equation}
acts as a projector onto the circular vectors,
\begin{equation} \label{projector}
\varepsilon_\mu(k,\lambda) = P_{\mu\nu}^{(\lambda)} e^{\nu(1)}\;.
\end{equation}
Eventually, we end up with a modified dispersion relation
\begin{equation}\label{DR}
G= (k^2)^2-4g^2(n \cdot k)^4 \left((n\cdot k)^2-n^2k^2\right)=0\;,
\end{equation}
and from the inverse of Eq.~\eqref{eq-mot} we arrive at the modified photon 
propagator in the axial gauge,
\begin{equation}\label{PROPAGATOR}
G_{ \mu \nu }(k) = -\sum_{\lambda} 
\frac{P^{(\lambda)}_{\mu \nu}(k)}{ k^2+2g\lambda (k\cdot n)^2\sqrt{D} }\;.
\end{equation}
\section{Unitarity}
With the preparations in the previous section we may
now compute discontinuities of amplitudes involving
the modified Lorentz-violating Lagrangian~\eqref{M-M-P}. 
Let us first focus on the forward Compton scattering at the one-loop
order with the corresponding Cutkosky cutting rules~\cite{Cut}, shown in Fig.~\ref{Fig1}. 
The imaginary part of the forward Compton scattering amplitude originates
from the discontinuities of the propagators in the loop -- along the cut on the
left-hand side of Fig.~\ref{Fig1}.
With respect to the Lagrangian~\eqref{M-M-P}
we now have to deal with a modified photon propagator~\eqref{PROPAGATOR}
as well as modified photon polarization vectors~\eqref{circular}.
Following the optical theorem, unitarity of this diagram requires
that 2 times the imaginary part of the forward scattering diagram 
$e^-(p_1) +\gamma(p_2) \rightarrow e^-(p_1) + \gamma(p_2)$ 
with the indicated loop is equal to the tree-level 
production cross section of photon plus electron, 
summed
over the two final physical polarization states (respectively spin states) and
integrated over the final-state phase space.
This identity is a direct consequence of the unitarity of the S-matrix.
In contrast to ordinary QED, here the modified photon propagator as
well as the modified polarization vectors of the photons change the
calculation of the loop amplitude and in principle could show a
violation of unitarity.

Nevertheless, we shall verify, 
that the left-hand side and the right-hand side of the diagrams shown
in Fig.~\ref{Fig1} are equal, and therefore they do fulfill unitarity.

The right-hand side can immediately be written down as
\begin{equation} \label{rhs}
\sum _{\lambda',s'}\int \frac{d^3 k}{(2\pi)^3 2 k_0} \frac{d^3 k'}{(2\pi)^32 k'_0}  
 \left| \mathcal M \right|^2 \delta^{(4)}
(k+k'-p_1-p_2)(2\pi)^4\;,
\end{equation}
with the matrix element
\begin{multline}
\mathcal M = -i e^2 
\frac{1}{(p^2-m^2)} \varepsilon^*_{\mu}(k,\lambda') \bar {u}(k',s') 
 \gamma^{\mu}(\sla {p}+m) \gamma^{\nu} 
\\ 
\times {u}(p_1,s)  \varepsilon_{\nu}(p_2,\lambda)\;,
\end{multline}
with $e$ the positron charge and $m$ the mass of the electron.
Note that the polarization vectors of the photons correspond to Eq.~\eqref{circular}
and the sum is over the two physical polarizations of the photon as well
as the two spin states of the fermion.
We now have to show that 2 times the imaginary part of the loop diagram, that is,
the left-hand-side of Fig.~\ref{Fig1}, equals Eq.~\eqref{rhs}.
\begin{figure}
\centering
\includegraphics[width=0.39\textwidth]{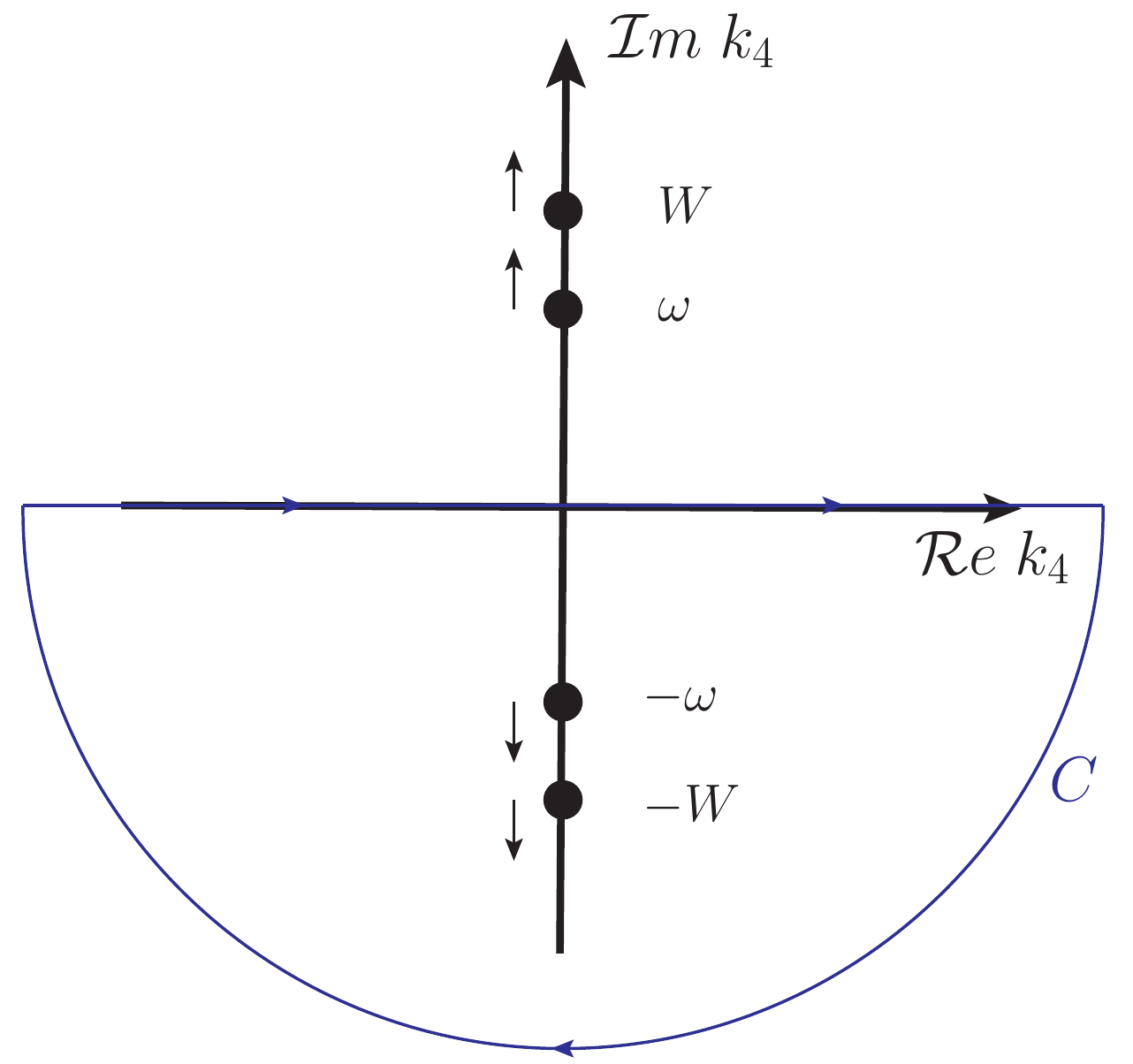}
\caption{\label{Fig2}The poles of the modified photon propagator in the 
complex $k_{4}$ plane and the 
 chosen contour $C$ of integration. The movement of the poles with
increasing energy is indicated by small arrows.}
\end{figure}
With the propagator (\ref{PROPAGATOR}) we get 
\begin{multline}
\mathcal A_\text{loop} = - \sum_{\lambda'}  e^4  
\frac{1}{(p^2-m^2)^2}\varepsilon^*_{\sigma}(p_2,\lambda)  \bar {u}(p_1,s) 
 \gamma^{\sigma}(\sla {p}+m)
 \gamma^{\mu} 
\\
 \times   \int \frac{d^4 q}{(2\pi)^4} 
\frac{(\sla {p}-\sla {q} +m)}{(p-q)^2-m^2}
\frac{\varepsilon_{\mu}(q,\lambda') \varepsilon^*_{\mu'}(q,\lambda') }{ q^2+2g\lambda'
 (q\cdot n)^2\sqrt{D} } 
\\
 \times  \gamma^{\mu'}  (\sla {p}+m ) \gamma^{\nu} u(p_1,s) 
\varepsilon_{\nu}(p_2,\lambda) \;,
\end{multline}
where we have replaced $P^{(\lambda)}_{\mu \nu}$ 
by employing Eq.~\eqref{imp-rel} in the propagator.
The polarization sum over $\lambda'$ now runs over all (real and virtual) 
polarizations of the
photon in the loop. By
renaming momenta, $k=q$, $k'=p-q$ and, using the identity
$\int \frac{d^4q}{(2\pi)^4}=\int \frac{d^4k'}{(2\pi)^4}  \int \frac{d^4k}{(2\pi)^4}
\delta^{(4)}(k+k'-p)(2\pi)^4$
we get 
\begin{multline}\label{Aloop}
\mathcal A_\text{loop} = - \sum_{\lambda'}  e^4  
\frac{1}{(p^2-m^2)^2} \varepsilon^*_{\sigma}(p_2,\lambda)  \bar {u}(p_1,s)
  \gamma^{\sigma}(\sla {p}+m)
 \gamma^{\mu} \\
\times   \int \frac{d^4 k'}{(2\pi)^4} 
\frac{(\sla {k}' +m)}{k'^2-m^2}\int \frac{d^4 k}{(2\pi)^4}
\delta^{(4)}(k+k'-p_1-p_2)(2\pi)^4
\\
\times \frac{\varepsilon_{\mu}(k,\lambda') \varepsilon^*_{\mu'}(k,\lambda')
 }{ k^2+2g\lambda'(k\cdot n)^2\sqrt{D} }  \gamma^{\mu'}  (\sla {p}+m ) 
\gamma^{\nu} u(p_1,s) 
\varepsilon_{\nu}(p_2,\lambda)\;.
\end{multline}
We now compute the imaginary part of $\mathcal A_\text{loop}$ by the computation of the
discontinuities of the loop propagators. 
The integration over $k_0$ can be performed by analytic continuation. 
We go to Euclidean
space with the replacement of the zero component of the four-vectors, 
for a generic four-vector $v$ , $v_{0}= -i v_{4}$ giving
the scalar product
$v v' = -v_1 v'_1 - v_2 v'_2 - v_3 v'_3 - v_4 v'_4 \equiv - v_\text{E} v'_\text{E}$. 
The propagator of the photon gives poles for a vanishing
denominator, which, in the Euclidean reads
\begin{multline}
 k_{E}^2 (-1+2ig   \lambda' \left| k_E \right| \left| n_E \right|^3 \left| \sin\theta \right|
\cos^2\theta  ) =
 \\
(k_4-\omega) 
(k_4+\omega) (-1+2ig   \lambda' \left| k_E \right| \left| n_E \right|^3 \left| \sin\theta \right|
\cos^2\theta  )\;.
\end{multline}
Hence, we encounter poles at $\pm \omega$ with
\begin{equation}
\omega  =  i| \vec k |\;,
\end{equation}
and two massive ghost modes at $\pm W$ with
\begin{equation}
W  = i \sqrt {| \vec k |^2 + M^2}\;, \quad 
M^2=\frac{1}{4g^2 n_E^6 \sin^2\theta \cos^4\theta} \;.
\end{equation}
These poles in the complex $k_4$ plane are shown in Fig.~\ref{Fig2}.
With increasing energy $k_0$ the poles move along the imaginary axis, 
departing from the origin, as indicated in the figure. 
In particular, they never cross the real $k_4$ axis.
We now close the integration contour along the real $k_4$ axis below and pick
up the residues in the lower semicircle. 
The residue at the massive mode at $k_4=-W$ will not occur in the phase-space integration
and therefore this residue does not contribute to the imaginary part.
Therefore we get for the imaginary part of the $k_4$ contour integration,
\begin{equation} \label{photonresidue}
 i \oint\limits_C \frac{ \varepsilon_{\mu}(k,\lambda') 
\varepsilon^*_{\mu'}(k,\lambda') d k_4}{(k_4-\omega) (k_4+\omega)
(-1+2ig   \lambda' \left| k_E \right| \left| n_E \right|^3 \left| \sin\theta \right|
\cos^2\theta  )}\;,
\end{equation}
only a contribution given by the residue at $k_4=-\omega=-i |\vec k|$, giving
\begin{equation}
i \cdot 2 \pi i  \lim_{k_4\rightarrow - \omega} 
\frac{\varepsilon_{\mu}(k,\lambda') \varepsilon^*_{\mu'}(k,\lambda')} {2 \omega} \;.
\end{equation}
We see that we only get a contribution to the discontinuity for on-shell photons.
Hence, the integration over $k_0$ in Eq.~\eqref{Aloop} singles out the
transverse polarization vectors of $\lambda'$ and gives a factor $i\pi/k_0$.

The integration over $k'_0$ can be performed as usual, where we
close the contour below in the Euclidean,
\begin{multline}
(-i) \oint\limits_C  \frac{ (\sla {k}' +m) d k'_4  }{  
(k'_4- i \sqrt{\vec k'^2 +m^2}) (k'_4+i \sqrt{\vec k'^2 +m^2})} =\\
(-i) 2\pi i \frac{(\sla {k}' +m)}{2 i k_0'} \;.
\end{multline}
Finally, with the spin-sum identity for on-shell fermions
$\sla{k}'+m = \sum _{s'}u(k',s')\bar{u}(k',s')$ we eventually arrive 
at the expression for the right-hand-side of Fig.~\ref{Fig1}, that is Eq.~\eqref{rhs}.
Once again, we stress, that the massive modes of the modified 
photon propagator do not contribute and unitarity is preserved for
the one-loop forward Compton scattering diagram.

Now, the step to one-loop unitarity of the Lorentz-violating model,
given by Eq.~\eqref{M-M-P}, follows directly: the only contributions
to the imaginary part of any one-loop diagram originate from the
discontinuities of the propagators, that is, from the poles of the 
propagators in the loops. In the case of a fermion propagator in the
loop, the poles give no additional imaginary contribution with respect to ordinary QED,
since the fermion part is unchanged.
In the case when a photon propagator appears in the loop, and hence,  may 
get on-shell in the loop integral, the integration
over the time-like component 
can always be written in the form~\eqref{photonresidue}. As we have shown, 
the massive modes of the propagator give no contribution to the imaginary
part in the contour integral.
Therefore, we conclude that unitarity is preserved at the one-loop order in
the model given by the Lorentz-violating Lagrangian~\eqref{M-M-P}.
\section{Conclusions}
Quantum gravity effects may be detected by the investigation
of Lorentz-violating terms in the Lagrangian, which are Planck-mass suppressed.
One higher-dimensional effective model is the Myers-Pospelev model
with a modified kinetic term of the photons.
In principle these modification could violate unitarity.
Here we have studied unitarity of this model at the one-loop order.
Firstly we recalled, how the photon sector has to be modified,
giving different photon polarization vectors and a different propagator.
We then studied unitarity in forward Compton-scattering at one loop and have
shown that unitarity is indeed preserved in this diagram.
In particular, we have shown that the additional massive modes of the photon propagator
do not contribute to the imaginary part of the loop diagram.
Finally, we have generalized this result
to an arbitrary one-loop Feynman diagram, that is,
the massive modes of the photon give no additional imaginary contributions.
This means that unitarity is preserved at the one-loop order.
\section*{Acknowledgments}
C.M.R acknowledges partial support 
from the Direcci\'on de Investigaci\'on de
la Universidad del B\'{\i}o-B\'{\i}o (DIUBB) Grant No. 123809 3/R
and FAPEI.

\end{document}